\documentclass[12pt]{article}
\usepackage[utf8x]{inputenc}
\usepackage[T1]{fontenc}

\usepackage{hfstyle}
\usepackage{graphicx}
\usepackage{multirow}
\usepackage[all,cmtip]{xy}
\usepackage{amsmath, amssymb,bbm,color}

\newtheorem{prop}{Proposition}

\newtheorem{lemma}{Lemma}

\newtheorem{conjecture}{Conjecture}

\newcommand{\NN}{{\mathbbm{N}}}

\newcommand{\B}{{{\mathcal{A}}^{\star}}}
\newcommand{\G}{{\mathcal A}}

\newcommand{\f}{{\mathbf f}}

\newcommand{\card}{\mathop{\mathrm{card}}}

\begin{document}
\title{Exponential convergence to equilibrium in cellular automata asymptotically emulating identity}
\author{Henryk Fuk\'s$^1$ { }and  Jos\'e Manuel G\'omez Soto$^2$ 
      \oneaddress{
  $^1$ Department of Mathematics,
   Brock University,
 St. Catharines, ON, Canada\\[1em]
         $^2$ Unidad Acad\'emica de Matem\'aticas, 
   Universidad Aut\'onoma de Zacatecas,
 Calzada Solidaridad entronque Paseo a la Bufa,
Zacatecas, Zac. M\'exico.\\[1em]
         \email{hfuks@brocku.ca, jmgomezgoo@gmail.com}
       }
   }

\Abstract{We consider the problem of finding the density of 1's in a configuration
obtained by $n$ iterations of a given cellular automaton (CA) rule, starting
from disordered initial condition. While this problems is intractable 
in full generality for a general CA rule, we argue that for some sufficiently
simple classes of rules it is possible to express the density in terms
of elementary functions. Rules asymptotically emulating identity
are one example of such a class, and density formulae have been
previously obtained for  several of them.
We show how to obtain formulae for density
for two further rules in this class, 160 and 168, and postulate likely   
expression for density for eight other rules. Our results are valid for arbitrary
initial density. Finally, we conjecture that the density
of 1's for CA rules asymptotically emulating identity
always approaches the equilibrium point exponentially fast.
}
%%% ----------------------------------------------------------------------
\maketitle
%%% --------

\section{Introduction}
Cellular automata (CA) are often viewed as computing devices.
An initial configuration is taken as an input of the computation, and,
after a number of iterations of the CA rule, the resulting final configuration
constitutes the output of the computation. 

In many practical problems, especially in mathematical modeling, 
one is not interested in all details of the configuration, but rather
in certain aggregate properties, such as, for example, the density of ones. 
A very common question can then be formulated as follows.
Suppose we generated an initial configuration with a given density of ones $p\in[0,1]$,
such that each site is independently set to 1 with probability $p$ and to
0 with probability $1-p$. We then  iterate a given rule $n$ times over this configuration. What is the density
of ones in the resulting configuration? 
Using signal processing terminology, we want to know  the ``response curve'',  density of 
the output as a function of the density of the input. 

Numerical studies of the density $c_n$ assuming $p=0.5$ were first conducted by S. Wolfram.
In \cite{Wolfram94}, he presented a table showing $c_{\infty}$ for all ``minimal'' CA rules,
 in many cases postulating exact rational values of $c_{\infty}$.
 In \cite{paper3}, H. Fuk\'s obtained formulae for density $c_n$ for many elementary
CA rules, starting from initial density $c_0=0.5$. Some of these formulae were proved, but most were conjectures
 based on patterns appearing in sequences of preimage numbers. 

In later years, building on the ideas outlined in \cite{paper3}, exact formulae
for $c_n$ have been rigorously derived for several CA rules, 
for example rules 14, 172, 140, and 130 \cite{paper34,paper39,paper44,paper40}. 
In the first two cases the forumulae for $c_n$ were proved for $p=1/2$,
while in the last two cases for arbitrary $p$.

For a given CA rule, the difficulty of finding the density $c_n$ very strongly depends on the rule.
Generally, the more complex dynamics of the rule is, the more difficult is to obtain the exact
formula for $c_n$. One exception to this are surjective CA rules (among elementary CA these are
rules 15, 30, 45, 51, 60, 90, 105, 106, 150, 154, 170 and 204). Some of them exhibit very complex
spatio-temporal behavior, yet  it is well known that the symmetric Bernoulli measure ($p=1/2$) is invariant
 under the action of a surjective rule, thus for all of them $c_n=1/2$ for $p=1/2$ (cf. 
\cite{Pivato2009} for a review of this result).

One class of rules for which $c_n$ is easy to obtain are idempotent rules, that is,
rules for which the global function $F$ has the property $F^2=F$ (rule applied twice yields
the same result as applied once). One can generalize the notion of idempotence further by
considering $k$-th level \emph{emulators of identity} , for which  $F^{k+1}=F^{k}$ for some $k$.
These are call emulators of identity, because after $k$ iterations further application of the rule is equivalent 
to application of the identity \cite{Rogers94}. And finally, one can introduce the notion 
of asymptotic emulation of identity, such that $F^{k+1}$ and $F^{k}$ are not identical, but
become closer and closer as $k \to \infty$, as defined in~\cite{paper3}. 
Rules asymptotically emulating identity will be the main subject of this paper.
While the dynamics of these rules  is not overly complicated, it is still far from being trivial.
In some sense, they resemble finitely-dimensional dynamical systems in the neighbourhood
of a hyperbolic fixed point, where orbits starting from the stable manifold converge to the fixed point exponentially
fast. In asymptotic emulators of identity, convergence to the equilibrium state is also exponentially fast, as we will
subsequently see.
For all the above reasons, CA rules asymptotically emulating identity are an ideal testbed for attempts to compute $c_n$. 
The goal of this article is to show that the problem of finding $c_n$ for these rules is indeed tractable,
and that their formulae for density exhibit  remarkable similarity to each other.

\section{Preliminaries and definitions}
Let ${\mathcal{A}}=\{0,1\}$ be called an \emph{alphabet}, or a \emph{symbol set}, 
and let $X={\mathcal{A}}^\mathbb{Z}$.
A finite sequence of elements of ${\mathcal{A}}$, $\mathbf{b}=b_1b_2\ldots, b_{n}$, will be called a \emph{block} 
 (or \emph{word})
 of length $n$.
Set of all blocks of elements of ${\mathcal{A}}$ of all possible lengths will be denoted by ${\mathcal{A}}^{\star}$.

For $r \in \NN$, a mapping $f:\G^{2r+1}\mapsto \G$ will be called {\em a cellular
 automaton rule of radius~$r$}. 
Corresponding to $f$, we also define a {\em global mapping}  $F:X \to X$ such that
$
(F(x))_i=f(x_{i-r},\ldots,x_i,\ldots,x_{i+r})
$
 for any $x\in X$.

A {\em block evolution operator} corresponding to $f$ is a mapping
 $\f:\B \mapsto \B$ defined as follows. 
Let $r\in \NN$ be the radius of $f$, and let  $\mathbf{a}=a_1a_2 \ldots a_{n}\in \G^n$
where $n \geq 2r+1$. Then $\f(\mathbf{a})$ is a block of length $n-2r$ defined as
\begin{equation}
\f(\mathbf{a}) = f(a_1,a_{2},\ldots,a_{1+2r})
f(a_2,a_{3},\ldots,a_{2+2r})\ldots
f(a_{n-2r},a_{n-2r+1},\ldots,a_{n}).
\end{equation}
For example, let $f$ be a rule of radius 1, 
and let $\mathbf{b}\in \mathcal{A}^5$, so that $\mathbf{b}=b_1b_2b_3b_4b_5$.
Then $\mathbf{f}(b_1b_2b_3b_4b_5)=a_1a_2a_3$, where $a_1=f(b_1,b_2,b_3)$, $a_2=f(b_2,b_3,b_4)$,
and $a_3=f(b_3,b_4,b_5)$. 
If  $\mathbf{f}(\mathbf{b})=\mathbf{a}$, than we will say that $\mathbf{b}$ is a preimage of
$\mathbf{a}$, and write $\mathbf{b} \in \mathbf{f}^{-1}(\mathbf{a})$.
Similarly, if $\mathbf{f}^n(\mathbf{b})=\mathbf{a}$, than we will say that $\mathbf{b}$ is an
\emph{$n$-step preimage} of
$\mathbf{a}$, and write $\mathbf{b} \in \mathbf{f}^{-n}(\mathbf{a})$.

The appropriate mathematical description of an initial distribution of
configurations is a probability measure $\mu$ on $X$ \cite{KurkaMaas2000,Kurka2005,Pivato2009,FormentiKurka2009}.  Suppose that
the initial distribution is a Bernoulli measure $\mu_{p}$, so 
and all sites are independently set to 1 or 0, and 
the probability of finding 1 at a given site is $p$ while the probability
of finding 0 is $1-p$. One can then show \cite{paper39} that the probability
$ P_n(\mathbf{b})$
of finding a block $\mathbf{b}$ at a given site after $n$ iterations of rule $f$ is given by
\begin{equation}
 P_n(\mathbf{b})= \sum_{\mathbf{a} \in \f^{-n}(\mathbf{b})}  P_0(\mathbf{a}).
\end{equation}
Note that the above probability is site-independent, and this is because the initial measure $\mu_p$ is shift-invariant.
We will define $c_n$, the density of 1's, to be the expected value of
a site, 
\begin{equation}
c_n= P_n(1)\cdot 1 + P_n(0) \cdot 0=P_n(1).
\end{equation}
This yields the expression for density
\begin{equation}
 c_n= \sum_{\mathbf{a} \in \f^{-n}(1)}  P_0(\mathbf{a}).
\end{equation}
Since the initial distribution is the Bernoulli distribution $\mu_p$, 
$P_0(\mathbf{a})=p^{\#_1 (\mathbf{a})} (1-p)^{\#_0 (\mathbf{a})}$, 
where $\#_1(\mathbf{a})$ and $\#_0 (\mathbf{a})$ denote, respectively,  the number of ones (zeros) in $\mathbf{b}$.
We then obtain
\begin{equation}
 c_n= \sum_{\mathbf{a} \in \f^{-n}(1)}  p^{\#_1 (\mathbf{a})} (1-p)^{\#_0 (\mathbf{a})}.
\end{equation}

In order to conveniently write the above formula, we will introduce the notion of a \emph{density polynomial}.
Let the \emph{density polynomial} associated with a binary string $\mathbf{b}=b_1b_2\ldots b_n$ be defined 
as 
\begin{equation}
 \Psi_{\mathbf{b}}(p,q)=p^{\#_1 (\mathbf{b})} q^{\#_0 (\mathbf{b})}.
\end{equation}
If $A$ is a set of binary strings, we define density polynomial associated with $A$ as
\begin{equation}
 \Psi_{A}(p,q)=\sum_{\mathbf{a} \in A} \Psi_{\mathbf{a}}(p,q).
\end{equation}
Density $c_n$ can thus be written as
\begin{equation}
 c_n=\Psi_{\f^{-n}(1)}(p,1-p)=\Psi_{\f^{-n}(1)}(c_0,1-c_0).
\end{equation}
In what follows, we will keep using variables
$p$ and $q$ for density polynomials, understanding that in order to obtain $c_n$,
 one needs to substitute $q=1-p$, and that $p$ is the initial density, $p=c_0$.

The problem of finding the density $c_n$ is thus equivalent tot he problem of finding the density
polynomial for the set $\f^{-n}(1)$. In order to do this, one has to have detailed knowledge of the 
structure of $\f^{-n}(1)$, which is usually very difficult to obtain. However, for
 reasonably simple rules it often possible, as we will shortly see.

\section{Asymptotic emulators of identity}
We will now define the class of rules we wish to consider, namely
rules asymptotically emulating identity.
Let $f$ be a CA rule of radius $m$, $g$ a rule of radius $n$, and $k=\max\{m,n\}$. Let the distance
between rules $f$ and $g$ be defined as
\begin{equation}
d(f,g)=2^{-2k-1}\sum_{\mathbf{b} \in {\mathcal{A}}^{2k+1}} \left|f(\mathbf{b})-g(\mathbf{b})\right|,
\end{equation}
where for $\mathbf{b}=b_1b_2\ldots b_{2k+1}$ and rule $f$ of radius $r$ we define $f(\mathbf{b})=
f(b_{k+1-r},\ldots,b_{k+1+r})$. This simply means that $f(\mathbf{b})$ is the value of 
the local function on the neighbourhood of the central symbol of $\mathbf{b}$, e.g.,
for $\mathbf{b}=b_1b_2b_3b_4b_5b_6b_7$ and $r=1$, $f(\mathbf{b})=f(b_3,b_4,b_5)$. One can show that the 
distance defined above is a metric in the space of CA rules \cite{paper3}.

The  \emph{composition} $f \circ g$ of two CA rules $f$ and  $g$ can be defined in terms
of their corresponding global mappings $F$ and $G$,  as a local function of $F\circ G$, where
$
(F\circ G)(x)=F(G(x))
$
for $x \in X$. We note that if  $f$ is a rule of radius $r$, and $g$ of radius $s$,
 then $f \circ g$ is a rule of radius $r+s$. For example, the composition of two
radius-1 mappings is a radius-2 mapping:
\begin{equation}
(f\circ g)(x_{-2},x_{-1},x_0,x_1,x_2)=
f(g(x_{-2},x_{-1},x_0),g(x_{-1},x_0,x_1),g(x_0,x_1,x_2)).
\end{equation} Multiple composition will be denoted by
\begin{equation}f^n=\underbrace{f \circ f
 \circ \cdots \circ f}_{\mbox{$n$ times}}.
\end{equation}
We say that a cellular automaton rule $f$ {\it asymptotically emulates
 rule $g$} if
\begin{equation}\lim_{n\to \infty}d(f^{n+1}, g\circ f^n)=0.
\end{equation}

We will be primarily interested in emulators of identity, for which we take as
$g$ the local function of identity rule (i.e., rule 204).  In \cite{paper3}, it has been
found that rules  13, 32, 40, 44,  77, 78, 128, 132, 136, 140, 160, 164, 168, 172, and 232 asymptotically
emulate identity. Typical spatio-temporal patterns produced by these rules are shown in
Figure 1. All  these rules eventually reach all zero state or a fixed point 
which corresponds to vertical strips in the spatio-temporal patters (as in the case of
rule 232, Figure 1d).
%%%%%%%%%%
\begin{figure}
(a)\includegraphics[scale=0.9]{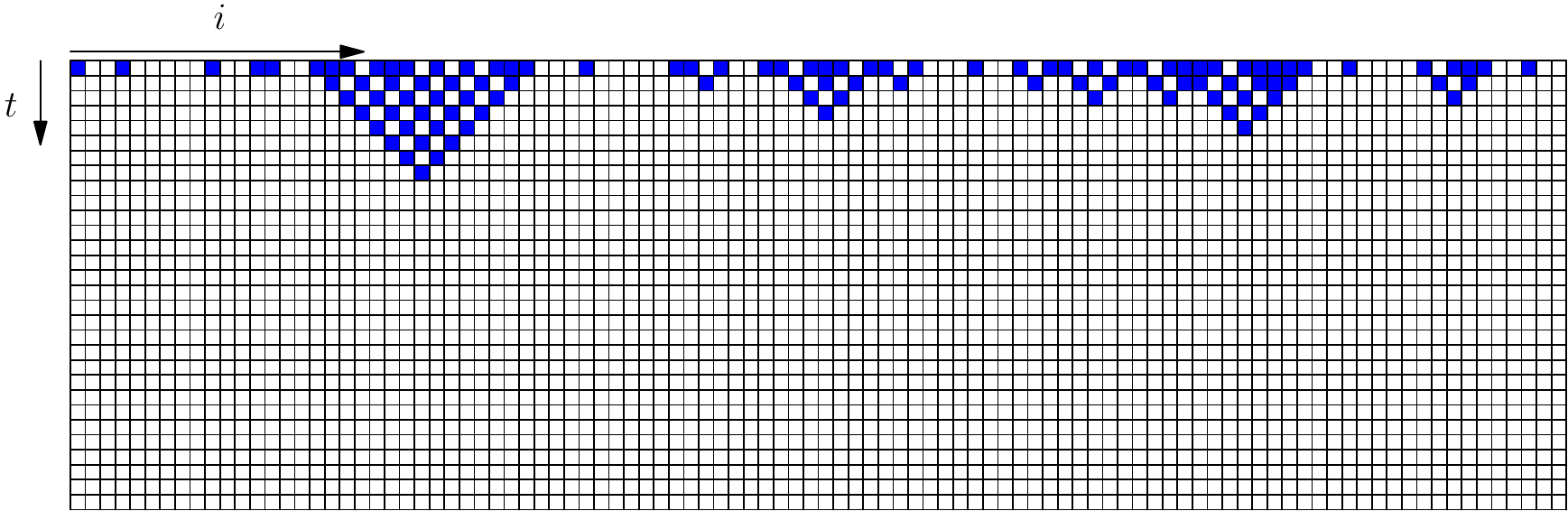}
(b)\includegraphics[scale=0.9]{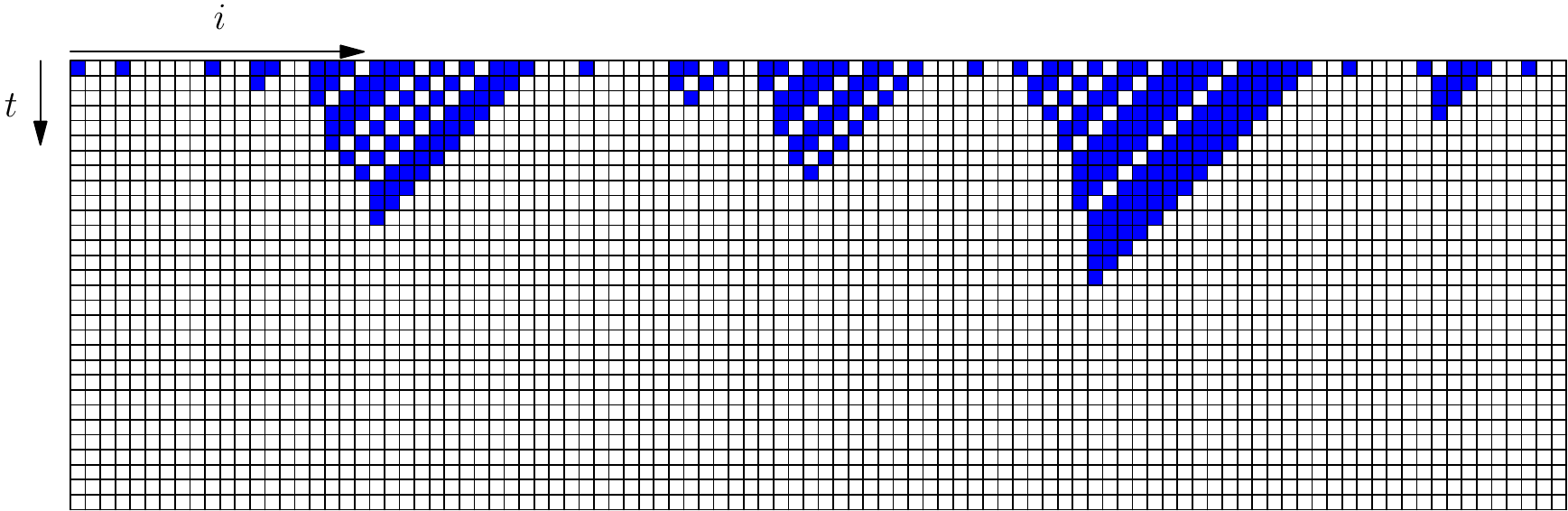}
(c)\includegraphics[scale=0.9]{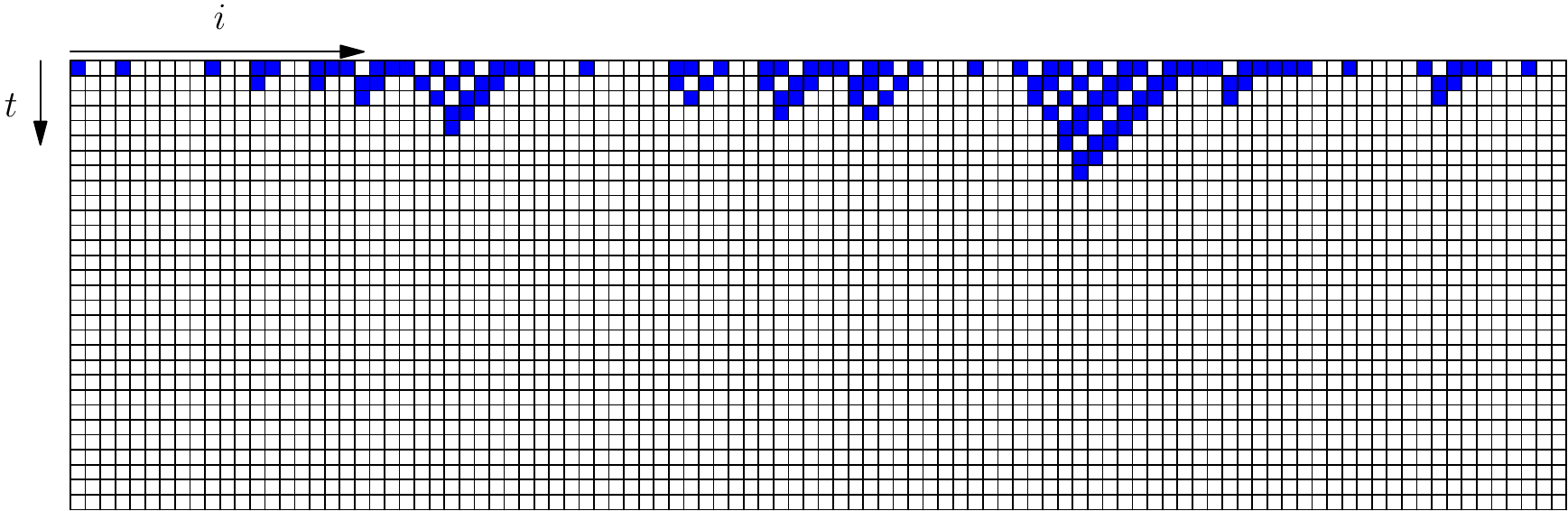}
(d)\includegraphics[scale=0.9]{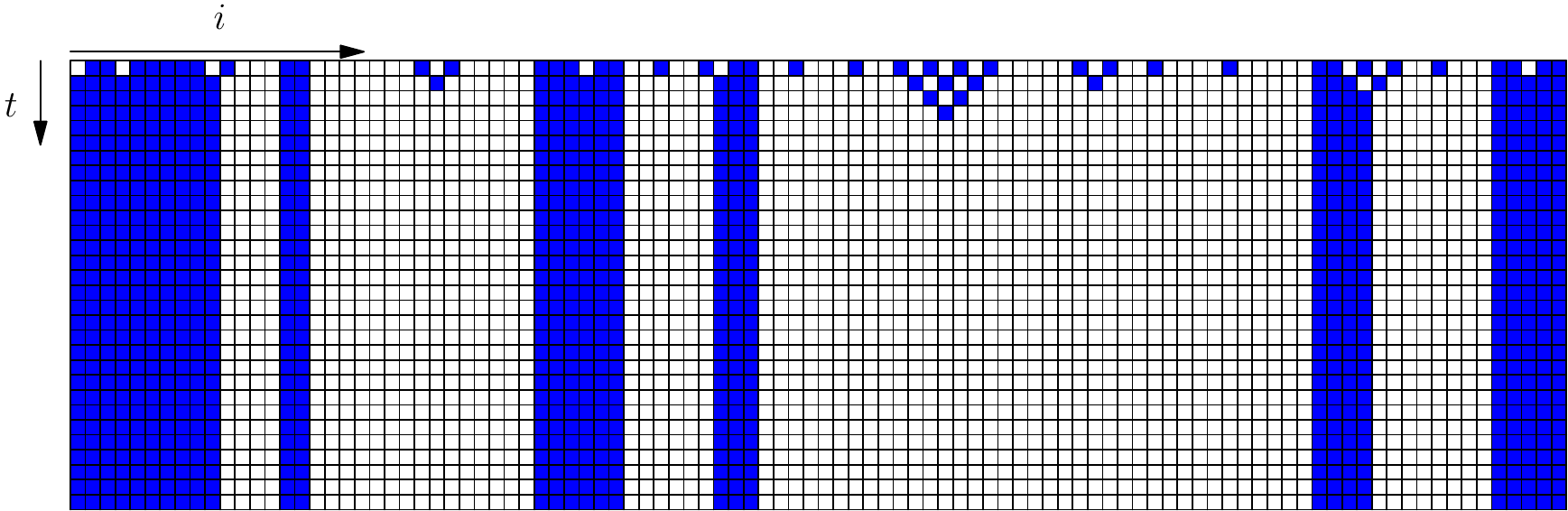}
\caption{Spatio-temporal pattern produced by rules  160 (a), 168 (b), 40 (c) , and 232 (d),
 starting with random initial condition.}
\end{figure}

For all these rules, formulae for densities for $c_n$ for $p=1/2$ have been postulated in \cite{paper3},
and some of these formulae were subsequently proved, as illustrated in Table 1.  The general formulae for the
density, for arbitrary $c_0$, have been previously reported for only four
of them, rules 128, 132, 136, and 140. For all four cases, proofs of the formulae are known.
Below we show these formulae, citing proof source as well.
\begin{itemize}
 \item  \textbf{Rule 128} (in \cite{paper7}, $c_n$ has been obtained for rule 254, identical with conjugated and reflected rule 128)
\begin{equation} \label{c128}
 c_n=c_0^{2n+1},
\end{equation}

  \item  \textbf{Rule 132} (in \cite{paper7}, $c_n$ has been obtained for rule 222, identical with conjugated and reflected rule 132)
\begin{equation} \label{c132}
 c_n=\left( 1-c_{{0}} \right) ^{2}c_{{0}}+{\frac { \left( 1-c_{{0}}
 \right) {c_{{0}}}^{3}}{1+c_{{0}}}}+2\,\frac {c_{{0}}}{1+c_{{0}}} {c_0}^{2 n+
1},
\end{equation}

\item  \textbf{Rule 136} (in \cite{paper7}, $c_n$ has been obtained for rule 238, identical with  conjugated rule 132)
\begin{equation} \label{c136}
 c_n=c_0^{n+1},
\end{equation}

 \item  \textbf{Rule 140} (in \cite{paper44}, $c_n$ has been obtained for a more general
 case of the asynchronous version rule 140, here we take the special case of
the  synchrony rate  equal to 1)
\begin{equation} \label{c140}
c_n= c_0(1-c_0)  + c_0^{n+2}.
\end{equation}
\end{itemize}

We will show 
that  using the concept of density polynomials, formulae for $c_n$ for arbitrary $c_0$ 
can be constructed for many other rules asymptotically emulating identity.
In two cases, namely for rules 160 and 168, we give formal proofs for density formulae.
For many other cases, we will describe how to ``guess''  the correct formula for $c_n$
by setting up a recursive equation for density polynomials.
%%%%%%%%%%
%%%%%%%%%
\begin{table}
\begin{center}
  \begin{tabular}{||c|l|c||}
\hline 
Rule  & $c_{n}$ & Proof \tabularnewline
\hline 
$13$ & $7/16-(-2)^{-n-3}$ &  \tabularnewline
\hline 
$32$
  & $2^{-1-2n}$ & \cite{paper3} \tabularnewline
\hline 
$40$  & $2^{-n-1}$ &  \tabularnewline
\hline 
$44$  & $1/6+\frac{5}{6}2^{-2n}$ &  \tabularnewline
\hline 
$77$  & $1/2$ & \cite{paper3} \tabularnewline
\hline 
$78$  & $9/16$ &  \tabularnewline
\hline 
$128$  & $2^{-1-2n}$ &  \cite{paper7}\tabularnewline
\hline 
$132$  & $1/6+\frac{1}{3}2^{-2n}$ & \cite{paper7} \tabularnewline
\hline 
$136$  & $2^{-n-1}$ &  \cite{paper7}\tabularnewline
\hline 
$140$  & $1/4+2^{-n-2}$ & \cite{paper44}   \tabularnewline
\hline 
$160$  & $2^{-n-1}$ & this paper \tabularnewline
\hline 
$164$  & $1/12-\frac{1}{3}4^{-n}+\frac{3}{4}2^{-n}$ &  \tabularnewline
\hline 
$168$  & $3^{n}2^{-2n-1}$ & this paper \tabularnewline
\hline 
$172$  & $\frac{1}{8}+\frac{(10-4\sqrt{5})(1-\sqrt{5})^{n}+(10+4\sqrt{5})(1+\sqrt{5})^{n}}{40\cdot2^{2n}}$ & \cite{paper39} \tabularnewline
\hline 
$232$  & $1/2$ &  \tabularnewline
\hline
\end{tabular}
\end{center}
\caption{Density of ones $c_n$ for disordered initial state  ($c_{0}=0.5$) for elementary rules
asymptotically emulating identity. For rules for which the proof
is known source of the proof is given. All others formulae are conjectures based on preimage patterns form \cite{paper3}.}
\end{table}

 \section{Rule 160}
The first rule we wish to consider is the rule 160. From now one, we will use subscripts with
Wolfram numbers to identify concrete local functions and corresponding block evolution operators,
{e.g., $f_{160}$ and $\mathbf{f}_{160}$ for rule 160.

Rule 160 is defined by $f_{160}(1,1,1)=f_{160}(1,0,1)=1$, and $f_{160}(x_1,x_2,x_3)=0$ for
all other values of $x_1,x_2,x_3$. This can be simply written as $f(x_1,x_2,x_3)=x_1x_3$.
Rule 160 is one of those few rules for which expressions for $f^n$ can be explicitly
given, as the following proposition attests.
\begin{prop}
 For elementary CA rule 160 and for any $n\in \NN$ we have
\begin{equation}\label{f160form}
 f^n_{160}(x_1,x_2,\ldots, x_{2n+1})=\prod_{i=0}^{n}x_{2i+1}.
\end{equation}
\end{prop}
\emph{Proof.} We give proof by induction. For $n=1$ eq. (\ref{f160form}) is obviously
true, as remarked above. Suppose now that the formula (\ref{f160form}) holds for some $n$,
and let us compute $f^{n+1}$. We have
\begin{align*}
 f^{n+1}_{160}(x_1,x_2,\ldots, x_{2n+3})&=
f_{160}\Big( f^{n}_{160}(x_1,\ldots, x_{2n+1}), 
f^{n}_{160}(x_2,\ldots, x_{2n+2}), 
f^{n}_{160}(x_3,\ldots, x_{2n+3}) \Big)\\
&=f_{160}\left( \prod_{i=0}^{n}x_{2i+1},
                \prod_{i=0}^{n}x_{2i+2},
                \prod_{i=0}^{n}x_{2i+3} 
  \right)=
\prod_{i=0}^{n}x_{2i+1} \prod_{i=0}^{n}x_{2i+3}\\
&=\prod_{i=0}^{n}x_{2i+1} \prod_{i=1}^{n+1}x_{2i+1}
=x_1\left(\prod_{i=1}^{n}x_{2i+1} \prod_{i=1}^{n}x_{2i+1}\right)x_{2n+3}
=\prod_{i=0}^{n+1}x_{2i+1},
\end{align*}
where we used the fact that $x_i^2=x_i$ if $x_i\in \{0,1\}$. The formula (\ref{f160form})
is thus valid for $n+1$, and this concludes the proof by induction.

The following result is a direct consequence of eq. (\ref{f160form}).
\begin{prop}
 Block $b_1b_2\ldots b_{2n+1}$ is an $n$-step preimage of 1 under the rule 160 if and only if
$b_i=1$ for every odd $i$.
\end{prop}
This means that we have $n+1$ ones and $n$ arbitrary symbols in the preimage of 1, 
  therefore,
\begin{equation}
\Psi_{\mathbf{f}^{-n}_{168}(1)}(p,q)=p^{n+1} (p+q)^n.
\end{equation}
The density of ones $c_n=P_n(1)$  is thus
\begin{equation} \label{c160}
c_n=\Psi_{\mathbf{f}^{-n}_{168}(1)}(c_0,1-c_0)=c_0^{n+1},
\end{equation}
and for $c_0=1/2$,
\begin{equation}
 c_n=2^{-n-1}.
\end{equation}
No matter what the initial density, $c_n$ exponentially converges to 0 as $n \to \infty$.
 \section{Rule 168}
Rule 168 is defined by $f_{168}(1,1,1)=f_{168}(1,0,1)=f_{168}(0,1,1)=1$, 
and $f_{168}(x_1,x_2,x_3)=0$ for all other values of $x_1,x_2,x_3$. Its dynamics
and preimage structure is considerably more complex that for rule 160.
Nevertheless, upon careful examination of preimages of 1, it is possible to
discover an interesting pattern in these preimages, described in the following
proposition.
\begin{prop}\label{prop168}
 Let $A_n$ be a set of all strings of length $2n+1$ ending with $1$ such that, counting from the right,
the first pair of zeros begins at $k$-th position from the right, and the number of isolated zeros 
in the substring to the right of this pair of zeros is $m$, satisfying $m<k-n-1$.
Moreover, let $B_n$ be the set of all strings of length $2n+1$ ending with $1$ which do not contain $00$.
Block $\mathbf{b} \in \mathcal{A}^{2n+1}$ is an $n$-step preimage of 1 under the rule 168 if and only if
$\mathbf{b} \in A_n \cup B_n$.
\end{prop}
In lieu of a formal proof, we will present discussion of spatio-temporal dynamics of rule
168 and explain how it leads to the above result. First of all, let us note that
$\mathbf{f}_{168}^{-1}(1)=\{011,101,111\}$. This means that if a block $\mathbf{b}$ ends with 1,
its preimage must also end with 1, and, by induction, its $n$-step preimage must end with 1 as well.
This explains that ending with 1 is a necessary condition for being a preimage of 1, and 
elements of both $A_n$ and $B_n$ have that property.

Next, let us note that one can consider a block $\mathbf{b}$ as consisting of blocks
of zeros of various lengths separated by blocks of ones of various length. Suppose that
 a given block contains one isolated zero and to the left of it a pair of adjacent zeros,
like in Figure~\ref{defects}.   
%%%%%%%%
\begin{figure}
\includegraphics[scale=0.8]{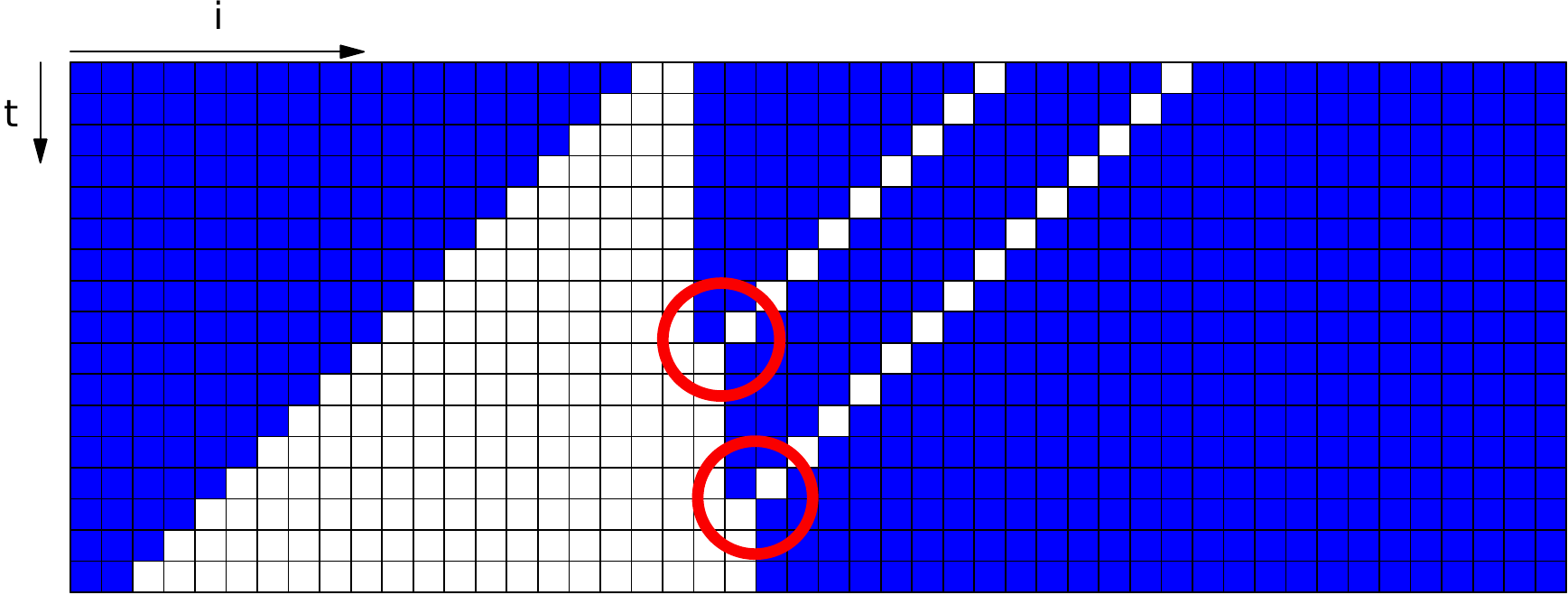}
\caption{Collision of ``defects'' in CA rule 168.}\label{defects}
\end{figure}
When the rule is iterated, the block 00 will increase its length by moving
its left boundary to the left, while its right boundary will remain in place.
The isolated zero, on the other hand, simply moves to the left, as illustrated
in Figure~\ref{defects}. When the boundary of the growing cluster of zeros collides
with the isolated zero, the isolated zero is annihilated, and the boundary of the 
cluster of zeros jumps one unit to the right. Two such collisions as shown 
in Figure~\ref{defects}, marked by circles.

Armed with this information, we can now attempt to describe conditions
which a block must satisfy in order to be an $n$-step preimage of 1. 
If a block of length $2n+1$ is an $n$-step preimage of 1, then either it contains a block of
two or more zeros or not. If it does not, and ends with 1, then it
necessarily is a preimage of 1. This is because when the rule is iterated, all
isolated zeros move to the left, and after $n$ iterations we obtain 1,
as shown in Figure~\ref{stringimages01} (left). Blocks of this type constitute
elements of $B_n$.

If, on the other hand, there is at least one  cluster of adjacent zeros 
in a block of length $2n+1$, then everything depends on the number of isolated zeros
to the right of the rightmost cluster of zeros. Clearly, if there are not
too many isolated zeros, and the rightmost cluster of zeros is not too far
to the right, then the collisions of isolated zeros with the boundary
of the cluster of zeros will not be able to move the boundary sufficiently
far to change the final outcome, which will remain 1. This situation
is illustrated in Figure~\ref{stringimages01} (center). Blocks of this
type are elements of $A_n$.

Obviously, the balance of clusters of zeros and individual zeros is
a delicate one, and if there are too many isolated zeros, they may
change the final outcome to 0, as in Figure~\ref{stringimages01} (right).

The question is then, what is the condition for this balance? 
To find this out, suppose that we have a string $\mathbf{b}\in \mathcal{A}^{2n+1}$
and  the first pair of zeros begins at $k$-th position from the right.
If there are no isolated zeros in the substring to the right of this pair, then
we want the end of the rightmost cluster of zeros  to be not further than just to the right
of the center of $\mathbf{b}$. Since the center of $\mathbf{b}$ is at the
$n+1$-th position from the right, we want  $k>n+1$. 

If  the are $m$ isolated zeros  in the substring to the right of this pair of zeros,
we must push the boundary of the rightmost cluster of zeros $m$ units
to the left, because these isolated zeros, after colliding with the rightmost cluster of zeros,
will move the boundary to the right. The condition should, therefore,
be in this case $k>n+1+m$, or, equivalently, $m<k-n-1$, as required for elements 
of $A_n$. $\square$
\begin{figure}
\begin{center}
 \begin{verbatim}
101101101101011101101     101111001111011101111     101111001111011101011
 1101101101011101101       1111000111011101111       1111000111011101011
  01101101011101101         11000011011101111         11000011011101011
   101101011101101           000001011101111           000001011101011
    1101011101101             0000011101111             0000011101011
     01011101101               00001101111               00001101011
      011101101                 000101111                 000101011
       1101101                   0001111                   0001011
        01101                     00111                     00011
         101                       011                       001
          1                         1                         0
\end{verbatim}
\end{center}
\caption{Examples of blocks of length 21 for which 10 iterations of $\mathbf{f}_{168}$ produce 1 (left and center) and 0 (right).}
 \label{stringimages01}
\end{figure}

With the above proposition, we can construct density polynomials associated with both $A_n$ and $B_n$.
%  For this, the following 
% lemma will be useful.
% \begin{lemma}\label{enumerationlemma}
% The number of binary strings
% $a_1a_2\ldots a_l$ such that $00$ does not appear as two consecutive terms $a_ia_{i+1}$ is equal to
% $F_{l+2}$, where $F_l$ is the $l$-th Fibonacci number.
% \end{lemma}
% From the lemma, we obtain 
% \begin{equation}
%  \card B_n=F_{2n +2}.
% \end{equation}
% For set $A_n$, let us use the following lemma.
The following lemma will be useful for this purpose. It can be proved by 
well known methods described in a typical book on enumerative combinatorics \cite{Stanley1986}.
 \begin{lemma}
The number of binary strings
$a_1a_2\ldots a_l$ such that $a_1=a_l=1$ and having only $m$ isolated zeros is
\begin{equation}
 \binom{l-m-1}{m}.
\end{equation}
 \end{lemma}
Now note that elements of the set $A_n$ described in Proposition~\ref{prop168} have the structure
\begin{equation}\label{structA}
 \underbrace{\star \ldots \star}_{2n-k} 00 a_1a_2 \ldots a_{k-1},
\end{equation}
where the string $a_1a_2 \ldots a_{k-1}$ has only isolated zeros and $a_1=a_{k-1}=1$.
Moreover, 
$$k \in \{n+2, n+3, \ldots , 2n\}.$$
 Furthermore, the number of isolated zeros $m$ must satisfy
$$m<k-n-1,$$
meaning that
\begin{equation}
 m \in \{ 0, 1, \ldots, k-n-2\}.
\end{equation}
Using our lemma, the density polynomial  of the set of strings of type (\ref{structA}) with fixed $k$ and $m$ is
therefore
\begin{equation}
 (p+q)^{2n-k}q^2 \binom{k-1-m-1}{m} q^m p^{k-m-1}= (p+q)^{2n-k} q^2 \binom{k-m-2}{m} q^m p^{k-m-1}.
\end{equation}
This yields the density polynomial associated with the set $A_n$,
\begin{equation}
\Psi_{A_n}(p,q)=\sum_{k=n+2}^{2n}  \sum_{m=0}^{k-n-2} (p+q)^{2n-k}  \binom{k-m-2}{m} q^{m+2} p^{k-m-1},
\end{equation}
which, by changing index $j$ to $k=n+j+2$, becomes
\begin{equation}
\Psi_{A_n}(p,q)=\sum_{j=0}^{n-2} \sum_{m=0}^{j} (p+q)^{n-j-2} \binom{n+j-m}{m}q^{m+2}p^{n+j-m+1}.
\end{equation}
For the set $B_n$, the associated density polynomial is
\begin{equation}
 \Psi_{B_n}(p,q)=\sum_{m=0}^n \binom{2n+1-m}{m}q^mp^{2n+1-m}.
\end{equation}
The resulting density polynomial for $n$-step preimages of 1 is, therefore,
\begin{align}
\Psi_{A_n \cup B_n}(p,q)=\Psi_{\mathbf{f}^{-n}_{168}(1)}(p,q)&=\sum_{j=0}^{n-2} \sum_{m=0}^{j} (p+q)^{n-j-2} \binom{n+j-m}{m}q^{m+2}p^{n+j-m+1}\\
+& \sum_{m=0}^n \binom{2n+1-m}{m}q^mp^{2n+1-m}.
\end{align}
This expression, while complicated, can be written in a closed form. 
One can namely show by induction (we omit the proof) that it sums to  
\begin{equation}
\Psi_{\mathbf{f}^{-n}_{168}(1)}(p,q)=p^{n+1}(p+2q)^{n}.
\end{equation}
If the initial density is $p=c_0$, $q=1-c_0$, we obtain
\begin{equation}\label{c168}
 c_n=\Psi_{\mathbf{f}^{-n}_{168}(1)}(c_0,1-c_0)=c_0^{n+1}(c_0+2-2c_0)^{n}=c_0^{n+1}(2-c_0)^{n}.
\end{equation}
For the symmetric case, $c_0=1/2$, 
\begin{equation}
 c_n=\Psi_{\mathbf{f}^{-n}_{168}(1)}(1/2,1/2)=\frac{3^n}{2^{2n+1}}.
\end{equation}
As in the case of rule 160, the density exponentially converges to 0 as $n \to \infty$.

As an interesting additional remark, note that by substituting $p=q=1$ to $\Psi_{\mathbf{f}^{-n}_{168}(1)}(p,q)$ 
we obtain $\card \mathbf{f}^{-n}_{168}(1)$,
thus 
\begin{equation}
 \card \mathbf{f}^{-n}_{168}(1)=\card A_n+\card B_n= \Psi_{\mathbf{f}^{-n}_{168}(1)}(1,1) =3^n.
\end{equation}
Density polynomials are thus useful not only for determining densities, but also to enumerate $n$-step
preimages in CA. The above result, $\card \mathbf{f}^{-n}_{168}(1) =3^n$, has been observed in \cite{paper3},
but no proof was given. 

\section{Rule 40}
In the previous two examples (rule 160 and 168), we were able to
gain detailed understanding of the structure of preimages of 1, and therefore
also compute the density of ones in a rigorous way.
In the next example this will not be the case, but  we will show that even 
then one can often conjecture what the expressions for $c_n$ are.
The conjecture will based on patters present  in density polynomials.
Such patters can often be readily observed when a first few
density polynomials are generated with the help of a computer program.

Let us now consider the rule 40, for which
$f_{40}(0,1,1)=f_{160}(1,0,1)=1$, and $f_{40}(x_1,x_2,x_3)=0$ for
all other values of $x_1,x_2,x_3$. The first 10 density polynomials for preimages of 1,
generated by a computer program,  are
\begin{align*}
\Psi_{\mathbf{f}^{-1}_{40}(1)}(p,q)= &2 p^2 q,\\
\Psi_{\mathbf{f}^{-2}_{40}(1)}(p,q)= &p^4 q+3 p^3 q^2,\\
\Psi_{\mathbf{f}^{-3}_{40}(1)}(p,q)= &3 p^5 q^2+5 p^4 q^3,\\
\Psi_{\mathbf{f}^{-4}_{40}(1)}(p,q)= &p^7 q^2+7 p^6 q^3+8 p^5 q^4,\\
\Psi_{\mathbf{f}^{-5}_{40}(1)}(p,q)= &4 p^8 q^3+15 p^7 q^4+13 p^6 q^5,\\
\Psi_{\mathbf{f}^{-6}_{40}(1)}(p,q)= &p^{10} q^3+12 p^9 q^4+30 p^8 q^5+21 p^7 q^6,\\
\Psi_{\mathbf{f}^{-7}_{40}(1)}(p,q)= &5 p^{11} q^4+31 p^{10} q^5+58 p^9 q^6+34 p^8 q^7,\\
\Psi_{\mathbf{f}^{-8}_{40}(1)}(p,q)= &p^{13} q^4+18 p^{12} q^5+73 p^{11} q^6+109 p^{10} q^7+55 p^9 q^8,\\
\Psi_{\mathbf{f}^{-9}_{40}(1)}(p,q)= &6 p^{14} q^5+54 p^{13} q^6+162 p^{12} q^7+201 p^{11} q^8+89 p^{10} q^9,\\
\Psi_{\mathbf{f}^{-10}_{40}(1)}(p,q)=&p^{16} q^5+25 p^{15} q^6+145 p^{14} q^7+344 p^{13} q^8+365 p^{12} q^9+144 p^{11} q^{10},\\
\Psi_{\mathbf{f}^{-11}_{40}(1)}(p,q)=&7 p^{17} q^6+85 p^{16} q^7+361 p^{15} q^8+707 p^{14} q^9+655 p^{13} q^{10}+233 p^{12} q^{11}.
\end{align*}
Upon closer inspection of these polynomials, one can suspect that they can perhaps be recursively generated.
Denoting for simplicity $U_n(p,q)=\Psi_{\mathbf{f}^{-n}_{40}(1)}(p,q)$, suppose that
they satisfy second-order difference equation,
\begin{equation}\label{receq}
 U_n(p,q)=\alpha(p,q)  U_{n-2}+ \beta(p,q) U_{n-1},
\end{equation}
where $\alpha(p,q)$ and $\beta(p,q)$ are some unknown functions. Polynomials
satisfying such a relation are known as generalized Lucas polynomials.

Knowing our first four polynomials, we can write the above equation for $n=3$ and $n=4$,
\begin{align}
 U_3(p,q)&=\alpha(p,q)  U_{1}+ \beta(p,q) U_{2},\nonumber \\
 U_4(p,q)&=\alpha(p,q)  U_{2}+ \beta(p,q) U_{3}.
\end{align}
This constitutes a system of two linear equations with two unknowns $\alpha(p,q)$ and $\beta(p,q)$.
Solving this system one obtains $\alpha(p,q)=p^2 q(p+q)$ and $\beta(p,q)=pq$,  
meaning that the recurrence equation (\ref{receq}) takes the form
\begin{equation}\label{r40locasrec}
 U_n(p,q)=p^2 q(p+q)  U_{n-2}+pq U_{n-1},
\end{equation}
where $U_0(p,q)=p$, $U_1(p,q)=2p^2q$. We verified that eq. (\ref{r40locasrec}) holds for
up to $n=12$, thus one can strongly suspect that it is valid for any $n$.

Assuming, therefore, that the linear difference equation (\ref{r40locasrec}) is valid for any $n$,
 we can now solve it by standard methods. The solution
is
\begin{multline}
U_n(p,q)=
-\frac{pq \left( -2\,p-q+\sqrt {5\,{q}^{2}+4\,pq} \right)}{\sqrt{5 q^2+4 p q} \left(q+\sqrt{5 q^2+4 p q}\right)}  \left( -{\frac {2
\,{p}^{2}q+2\,p{q}^{2}}{q+\sqrt {5\,{q}^{2}+4\,pq}}} \right) ^{n}\\
-\frac{pq \left( 2\,p+q+\sqrt {5\,{q}^{2}+4\,pq} \right)}{{\sqrt {5\,{q}^{2}+4\,pq} \left( q-\sqrt {5\,
{q}^{2}+4\,pq} \right)  }} 
 \left( -{\frac {2\,{p}^{2}q+2\,p{q}^{2}}{q-\sqrt {5\,{q}^{2}+4\,pq}}}
 \right) ^{n}.
\end{multline}
The density $c_n$ can now be computed by taking $c_n=U_n(c_0,1-c_0)$, after simplification and rationalization
yielding
% \begin{align}
% c_n=
% -&\frac{c_0(1-c_0) \left( -c_0-1+\sqrt {(c_0-1)(c_0-5)} \right)}{\sqrt{(c_0-1)(c_0-5)} 
% \left(1-c_0+\sqrt{(c_0-1)(c_0-5)}\right)}  \left( {\frac {2c_0(c_0-1)}
% {q+\sqrt {(c_0-1)(c_0-5)}}} \right) ^{n}\nonumber\\
% -&\frac{c_0(1-c_0) \left( c_0+1+\sqrt {(c_0-1)(c_0-5)} \right)}{{\sqrt {(c_0-1)(c_0-5)} \left( 1-c_0-\sqrt {(c_0-1)(c_0-5)} 
% \right)  }} 
%  \left( {\frac {2c_0(c_0-1)}{q-\sqrt {(c_0-1)(c_0-5)}}}
%  \right) ^{n}.
% \end{align}
\begin{multline} \label{c40}
 c_n= \left( \frac{1}{2}\,c_{{0}}-\frac{3}{2}\,{\frac {c_{{0}}\sqrt {5-6\,c_{{0}}+{c_{{0}}}
^{2}}}{c_{{0}}-5}} \right)  \left( \frac{1}{2}\, \left( 1-c_{{0}}+\sqrt {5-6\,
c_{{0}}+{c_{{0}}}^{2}} \right) c_{{0}} \right) ^{n} \\
+ \left( \frac{1}{2}\,c_{{0
}}+\frac{3}{2}\,{\frac {c_{{0}}\sqrt {5-6\,c_{{0}}+{c_{{0}}}^{2}}}{c_{{0}}-5}}
 \right)  \left( \frac{1}{2}\, \left( 1-c_{{0}}-\sqrt {5-6\,c_{{0}}+{c_{{0}}}^
{2}} \right) c_{{0}} \right) ^{n}.
\end{multline}

In the symmetric case $c_0=1/2$ we obtain, after simplification,
\begin{equation} \label{denssym40}
 c_n=2^{-n-1}.  
\end{equation}

For the symmetric case $c_0=1/2$, it is also possible to obtain the above expression for $c_n$  by a different 
method. One can show (we omit the proof here) that the generalized Lucas polynomials $U_n(p,q)$ 
defined by eq. (\ref{r40locasrec})
can be written in the form
\begin{equation}
 U_n(p,q)=\Psi_{\mathbf{f}^{-n}_{40}(1)}(p,q)=\sum_{k=1}^{n+1} T_{n+1,k} p^{2n+2-k} q^{k-1},  
\end{equation}
where the values of $T_{n,k}$ form the triangle
\begin{gather*}                                                      
                                   0, 2\\
                                 0, 1, 3\\
                                0, 0, 3, 5\\
                              0, 0, 1, 7, 8\\
                            0, 0, 0, 4, 15, 13\\
                          0, 0, 0, 1, 12, 30, 21\\
                        0, 0, 0, 0, 5, 31, 58, 34\\
                      0, 0, 0, 0, 1, 18, 73, 109, 55\\
                    0, 0, 0, 0, 0, 6, 54, 162, 201, 89\\
                 0, 0, 0, 0, 0, 1, 25, 145, 344, 365, 144.
\end{gather*}
The above triangle is known as the skew triangle associated with the Fibonacci numbers~\cite{A084938}.
The coefficients $T_{n,k}$ can be generated by the recursive procedure \cite{A084938},
%OEIS Foundation Inc. (2011), The On-Line Encyclopedia of Integer Sequences, http://oeis.org/A084938.
\begin{align}
 T_{n,k}&=T_{n-1,k-1}+T_{n-2,k-1}+T_{n-2,k-2},\\
 T_{n,k}&=0 \mbox{\,\,\,if\,\,\,} k<0  \mbox{\,\,\,or\,\,\,} k>n, \nonumber \\
 T_{0,0}&=1, \mbox{\,\,\,\,\,} T_{2,1}=0. \nonumber
\end{align}
Let us now compute $c_n$ for the symmetric initial condition $c_0=1/2$,
\begin{equation}
 c_n=\Psi_{\mathbf{f}^{-n}_{40}(1)}(1/2,1/2)=2^{-2n-1}\sum_{k=1}^{n+1} T_{n+1,k}.  
\end{equation}
Define now 
\begin{equation}
S_n= \sum_{k=1}^{n} T_{n,k},
\end{equation}
so that
\begin{equation}
 c_n=2^{-2n-1}S_{n+1}.  
\end{equation}
Using the recursion definition of $T$, we obtain
\begin{equation}
 \sum_{k=1}^{n}T_{n,k}=\sum_{k=1}^{n}T_{n-1,k-1}+
\sum_{k=1}^{n}T_{n-2,k-1}+\sum_{k=1}^{n+1}T_{n-2,k-2},
\end{equation}
hence
\begin{equation}
 S_{n}=S_{n-1}+2S_{n-2}.
\end{equation}
From the definition of $T(n,k)$ we know that $S_1=1$ and $S_2=2$, and therefore the solution 
of the above second-order difference equation is $S_n=2^n$, hence
\begin{equation}
 c_n=2^{-2n-1}\cdot 2^n=2^{-n-1},  
\end{equation}
the same as in eq. (\ref{denssym40}), as expected.

\section{Rules 232, 13, 32, 77, 78, 172, and 44}
Elementary CA rule 232 is a special case of the ``majority voting rule'' with radius 1,
defined as 
\begin{equation}
 f_{232}(x_1,x_2,x_3)=\mathrm{majority}\{x_1,x_2,x_3\},
\end{equation}
or, more explicitly, $f_{232}(1,1,1)=f_{232}(1,1,0)=f_{232}(1,0,1)=f_{232}(0,1,1)=1$, 
and for all other values of $x_1,x_2,x_3$,  $f_{232}(x_1,x_2,x_3)=0$.

We proceed in a similar fashion as in the case of rule 40.
The first few density polynomials are
\begin{align*}
\Psi_{\mathbf{f}^{-1}_{40}(1)}(p,q)= &3\,q{p}^{2}+{p}^{3},\\
\Psi_{\mathbf{f}^{-2}_{40}(1)}(p,q)= &{p}^{5}+5\,{p}^{4}q+8\,{p}^{3}{q}^{2}+2\,{p}^{2}{q}^{3},\\
\Psi_{\mathbf{f}^{-3}_{40}(1)}(p,q)= &{p}^{7}+7\,{p}^{6}q+19\,{p}^{5}{q}^{2}+24\,{p}^{4}{q}^{3}+11\,{p}^{3}{
q}^{4}+2\,{p}^{2}{q}^{5},\\
\Psi_{\mathbf{f}^{-4}_{40}(1)}(p,q)= &{p}^{9}+9\,{p}^{8}q+34\,{p}^{7}{q}^{2}+69\,{p}^{6}{q}^{3}+79\,{p}^{5}{
q}^{4}+47\,{p}^{4}{q}^{5}+15\,{p}^{3}{q}^{6}+2\,{p}^{2}{q}^{7},
\\
\Psi_{\mathbf{f}^{-5}_{40}(1)}(p,q)= &{p}^{11}+11\,{p}^{10}q+53\,{p}^{9}{q}^{2}+146\,{p}^{8}{q}^{3}
+251\,{p}
^{7}{q}^{4}\\
&\mbox{\hspace{10em}}+275\,{p}^{6}{q}^{5}+187\,{p}^{5}{q}^{6}+79\,{p}^{4}{q}^{7}
+19\,{p}^{3}{q}^{8}+2\,{p}^{2}{q}^{9},\\
\ldots &,
\end{align*}
and again, upon closer inspection it turns out that that they are generalized Lucas polynomials.
Denoting  $U_n(p,q)=\Psi_{\mathbf{f}^{-n}_{132}(1)}(p,q)$,
these polynomials satisfy
\begin{equation}\label{r232locasrec}
 U_n(p,q)=-pq(p+q)^2 U_{n-2}(p,q)+(p^2+3 pq+q^2)U_{n-1}(p,q).
\end{equation}
Solution of the above equation is
\begin{equation}
U_n(p,q)={\frac {{p}^{2} \left( p+2\,q \right)  \left( {p}^{2}+2\,pq+{q}^{2}
 \right) ^{n}}{{p}^{2}+pq+{q}^{2}}}-{\frac { \left( p-q \right) 
 \left( pq \right) ^{n+1}}{{p}^{2}+pq+{q}^{2}}}.
\end{equation}
The density $c_n$ can now be computed by taking $c_n=U_n(c_0,1-c_0)$, yielding
\begin{equation}\label{c232}
c_n= {\frac {{c_0}^{2} \left(2 -c_0 \right) }{{c_0}^{2}-c_0+1}}+{\frac { \left( 2
\,c_0-1 \right) c_0 \left( c_0-1 \right)  \left( c_0 \left( 1-c_0 \right) 
 \right) ^{n}}{{c_0}^{2}-c_0+1}}.
\end{equation}
We can see that $c_n$ exponentially converges to $c_{\infty}$,
where
\begin{equation}
 c_{\infty}={\frac {{c_0}^{2} \left(2 -c_0 \right) }{{c_0}^{2}-c_0+1}}.
\end{equation}

For $c_0=1/2$, the second term in eq. (\ref{c232})  vanishes and $c_{\infty}=1/2$, thus
we obtain $c_n=1/2$, in agreement with Table 1.

There are six other rules for which we were able to obtain expressions for
$c_n$ in the same way as above, except that the order of the difference equation
for density polynomials was not always 2, like in eq. (\ref{r232locasrec}), but
it was sometimes lower or (most of the time) higher. 
For these rules, which are 13, 32, 77, 78, 172, and 44,
we give below the recurrence formula for the density polynomial,
followed by the expression for $c_n$ obtained by solving that recurrence equation.
\begin{itemize}
\item \textbf{Rule 13:}
\begin{equation}
 U_n (p,q) = qp \left( q+p \right) ^{4} U_{n-3} (p,q) +
 \left( {q}^{2}+pq+{p}^{2} \right)  \left( q+p \right) ^{2}U_{n-2}(p,q), 
\end{equation}
%%%%%%%%%%
\begin{multline} \label{c13}
 c_n={\frac { \left( 1-c_{{0}} \right) ^{3} \left( -1+c_{{0}} \right) ^{n}}
{c_{{0}}-2}}+{\frac {{c_{{0}}}^{2} \left( {c_{{0}}}^{2}-2\,c_{{0}}+2
 \right)  \left( -c_{{0}} \right) ^{n}}{c_{{0}}+1}}
+{\frac { \left( {c
_{{0}}}^{3}-2\,{c_{{0}}}^{2}+c_{{0}} \right)^2-1   }{ \left( c_{{0}}-2 \right)  \left( 
c_{{0}}+1 \right) }}.
\end{multline}
\item \textbf{Rule 32:}
\begin{equation}
 U_n(p,q)=pq U_{n-1}(p,q),
\end{equation}
\begin{equation} \label{c32}
 c_n=c_0^{n+1}(1-c_0)^n.
\end{equation}

\item  \textbf{Rule 77:}
\begin{multline}
 U_n (p,q) = \left( p+q \right) ^{2}{q}^{2}{p}^{2}U_{n-3}(p,q) \\
+ \left( {p}^{4}+2 q{p}^{3}+{q}^{2}{p}^{2}+2\,{q}^{3}p+{q}^{
4} \right) U_{n-2}(p,q) +2 p q U_{n-1} (p,q),
\end{multline}
\begin{multline} \label{c77}
c_n= {\frac {{c_{{0}}}^{3} \left( -{c_{{0}}}^{2} \right) ^{n}}{{c_{{0}}}^{2
}+1}}
-{\frac { \left( 1-c_{{0}} \right) ^{3}
 \left( - \left( 1-c_{{0}} \right) ^{2} \right) ^{n}}{{c_{{0}}}^{2}-2
\,c_{{0}}+2}}
-{\frac {{c_{{0}}}^{5}-3\,{c_{{0}}}^{4}+3\,{c_{{0}}}^{3}-2\,{c_{{0
}}}^{2}+c_{{0}}-1}{ \left( {c_{{0}}}^{2}+1 \right)  \left( {c_{{0}}}^{
2}-2\,c_{{0}}+2 \right) }}.
\end{multline}

\item \textbf{Rule 78:}
\begin{multline}
U_{n}(p,q) = (p+q)^6 q^2 p^2 U_{n-5}(p,q)-(p+q)^4 q^2 p^2 U_{n-4}(p,q)
-(p^2+q^2) (p+q)^4 U_{n-3}(p,q)\\
+(p^2+q^2) (p+q)^2 U_{n-2}(p,q)+(p+q)^2 U_{n-1}(p,q),
\end{multline}
\begin{multline} \label{c78}
 c_n={\frac {1+c_{{0}}-{c_{{0}}}^{2}+{c_{{0}}}^{4}-2\,{c_{{0}}}^{5}+{c_{{0}
}}^{6}}{ \left( c_{{0}}+1 \right)  \left( 2-c_{{0}} \right) }}+\frac{1}{2}\,{
\frac { \left( 2\,{c_{{0}}}^{2}+1-2\,c_{{0}} \right) c_{{0}} \left( 1-
c_{{0}} \right)  \left( c_{{0}}-1 \right) ^{n}}{2-c_{{0}}}}\\
-\frac{1}{2}\,
 \left( 2\,c_{{0}}-1 \right) {c_{{0}}}^{2}{c_{{0}}}^{n}-\frac{1}{2}\,{\frac {
 \left( 1-c_{{0}} \right) {c_{{0}}}^{2} \left( -c_{{0}} \right) ^{n}}{
c_{{0}}+1}}+\frac{1}{2}\, \left( 1-c_{{0}} \right)  \left( 2\,c_{{0}}-1
 \right)  \left( 1-c_{{0}} \right) ^{n}.
\end{multline}
The above is valid for $n>1$.

\item \textbf{Rule 172:}
\begin{equation}
 U_n(p,q) =- p q \left( q+p \right) ^{4}U_{n-3}(p,q) -
 \left( q+p \right) ^{2}{p}^{2}U_{n-2}( p, q ) + \left( q+p
 \right)  \left( q+2\,p \right) U_{n-1}(p,q ),
\end{equation}
%%%%%%%%%%%%%
\begin{multline} \label{c172}
c_n=\left(c_{{0}}-1 \right) ^{2}c_{{0}}\\
-{\frac { \left( 3\,c_{{0}}-4+
\sqrt {4 c_{0} -3 c_0^2} \right)  \left( c_{{0}}-2+\sqrt {
4 c_{0} -3 c_0^2} \right) c_{{0}} \left( \frac{1}{2}\,c_{{0}}-\frac{1}{2}
\,\sqrt {4 c_{0} -3 c_0^2} \right) ^{n}}{12\,c_{{0}}-16}}\\
+{\frac { \left( 3\,c_{{0}}-4-\sqrt {4 c_{0} -3 c_0^2}
 \right)  \left( -c_{{0}}+2+\sqrt {4 c_{0} -3 c_0^2}
 \right) c_{{0}} \left( \frac{1}{2}\,c_{{0}}
+\frac{1}{2}\,\sqrt {4 c_{0} -3 c_0^2} \right) ^{n}}{12\,c_{{0}}-16}}.
\end{multline}

\item \textbf{Rule 44:}
\begin{equation}
 U_{n}(p,q) = -(p+q)^2 q^2 p^4 U_{n-4}(p,q)+q^2 p^4 U_{n-3}(p,q)+(p+q)^2 U_{n-1}(p,q),
\end{equation}
%%%%%%%%%%%%%%%%%%
\begin{multline} \label{c44}
 c_n={\frac { \left( {c_{{0}}}^{2}-c_{{0}}+1 \right) c_{{0}} 
\left( c_{{0}}-1 \right)
 }{{c_{{0}}}^{3}-{c_{{0}}}^{2}-1}}
-\frac{1}{3}\,{\frac {c_{{0}}}{1
+{c_{{0}}}^{2} \left( 1-c_{{0}} \right) }} \Big(\alpha \lambda_1^n + (\beta + i\, \gamma)\lambda_2^n+
(\beta - i\, \gamma)\lambda_3^n \Big),
\end{multline}
where
\begin{align*}
\lambda_1= {c_{{0}}}^{4/3} \left( 1-c_{{0}} \right) ^{2/3},\,\,\,\,\,\,\,
\lambda_{2,3}=\mp\frac{1}{2}\,{c_{{0}}}^{4/3} \left( 1-c_{{0}} \right) ^{2/3} \left(\pm 1+i\sqrt 
{3} \right),
\end{align*}
and
\begin{align*}
%%%%%%%
 \alpha&=- \left( 1+c_{{0}} \right)  \left( 1+c_{{0}}-{c_{{0}}}^{2} \right) 
-{\frac {\sqrt [3]{1-c_{{0}}}}{{c_{{0}}}^{2/3}}} \Delta,\\
 \beta&=- \left( 1+c_{{0}} \right)  \left( 1+c_{{0}}-{c_{{0}}}^{2} \right) 
+\frac{1}{2}\,{\frac {\sqrt [3]{1-c_{{0}}}}{{c_{{0}}}^{2/3}}} \Delta,\\
\gamma&= -\frac{\sqrt {3}}{2}\,{\frac {\sqrt[3]{1-c_{{0}}} \left( \Delta-2\,\sqrt [3]
{1-c_{{0}}} \left( 2-c_{{0}} \right)  \left( 1+{c_{{0}}}^{2} \right) 
 \right) }{{c_{{0}}}^{2/3}}},\\
%%%%%
\Delta&=\sqrt [3]{c_{{0}}} \left( 2-{c_{{0}}}^{3}\right) -\sqrt [3]{1-c_{{0}
}} \left( c_{{0}}-2 \right)  \left( 1+{c_{{0}}}^{2} \right). 
\end{align*}
\end{itemize}
\section{The remaining rule}
Among 15 CA rules asymptotically emulating identity, 
we either proved or conjectured general expressions for $c_n$ for 14 of them. In all cases,
exponential convergence to $c_\infty$ can be observed. What remains is only rule the 164 for 
which we were not able to find a closed form expression for density polynomials. 
We have attempted to find recurrence equations up to $6$-th order for 
this rule, to no avail. One suspects that the reason for this is dynamics
of rule 164, far more complicated than for other rule considered in this paper.
In Figure \ref{dens164}(a), one can clearly see that in the spatio-temporal
pattern generated by this rule exhibits the characteristic triangles of varying size.
Similar triangles are frequently observed in complex ``chaotic''
rules. 

Nevertheless, we have studied behaviour of $c_n$ numerically.
Figure \ref{dens164}(b) shows semi-logarithmic plots of $|c_n - c_{\infty}|$ as a function of $n$,
obtained by averaging 100 runs of simulations using a lattice with
$10^5$ sites. The value of $c_{\infty}$ in each case has been taken as the steady-state
value, that is, the final value of $c_n$ which was no longer changing. 
\begin{figure}
\begin{center}
(a)\includegraphics[scale=0.8]{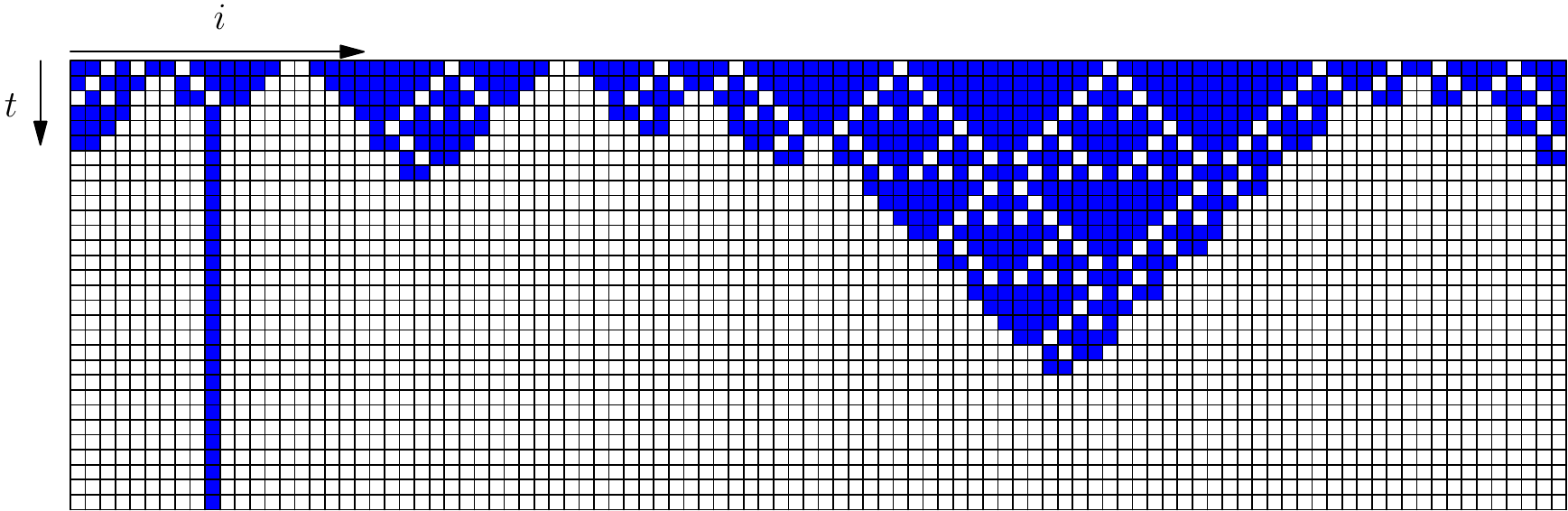}
(b)\includegraphics[scale=1.0]{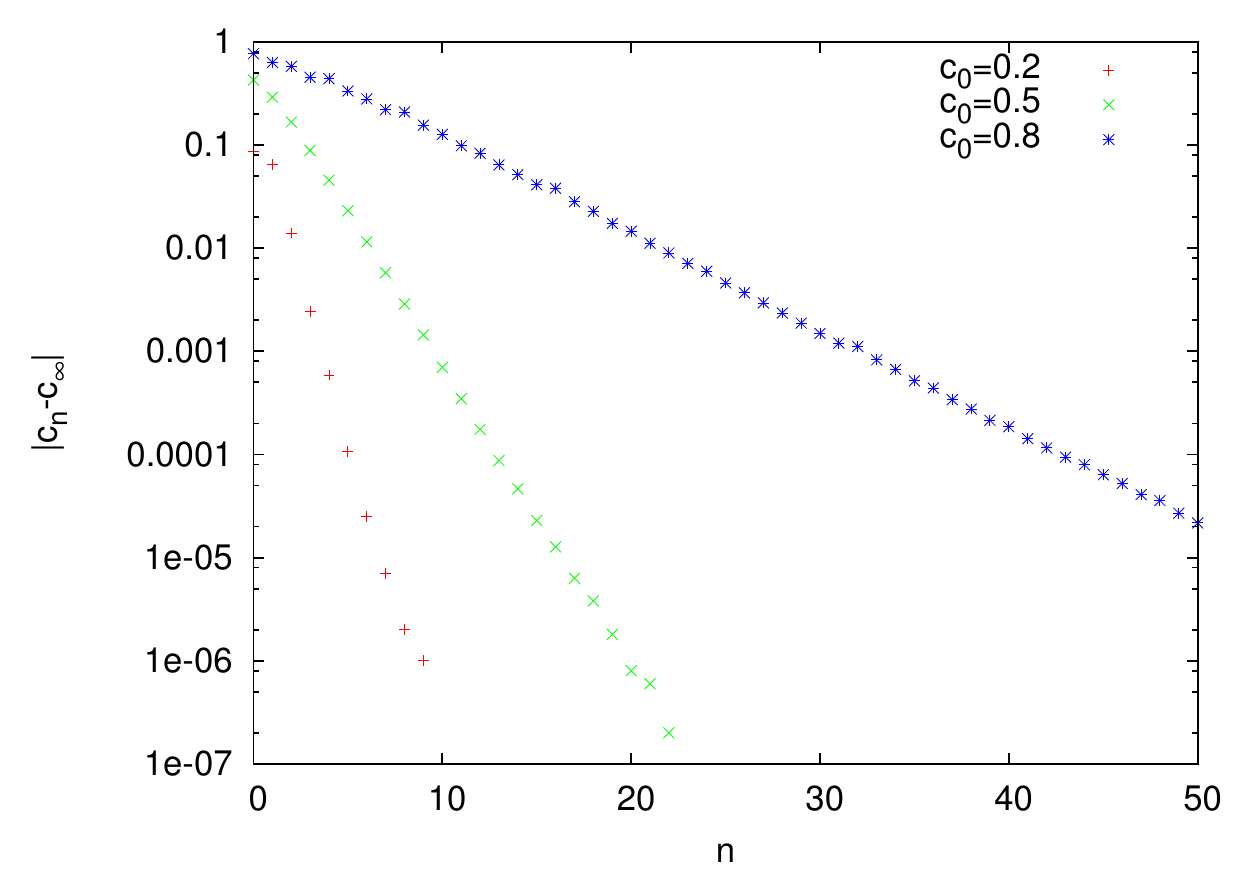}
\end{center}
\caption{ (a) Spatio-temporal pattern for rule 164, starting from random
initial condition with density 0.85. (b) Density $c_n$ as a function of $n$ for rule 164.
Lattice with $10^5$ sites and periodic configurations was used. Each points
corresponds to average of 100 experiments.} \label{dens164}
\end{figure}
From this plots it is clear that the graphs of $|c_n - c_{\infty}|$ vs. $n$
closely follow straight lines in all cases, strongly suggesting that the approach to the 
fixed point is also exponential, just like for the other 14 rules.

\section{Conclusions}

\begin{table}[t]
\begin{center}
  \begin{tabular}{||c|l|c||}
\hline 
Rule  & $c_{n}$ & Proof/conjecture \tabularnewline
\hline 
$13$ & eq. (\ref{c13}) & conj.  \tabularnewline
\hline 
$32$  & eq. (\ref{c32})  & conj.  \tabularnewline
\hline 
$40$  & eq. (\ref{c40}) &  conj.  \tabularnewline
\hline 
$44$  & eq. (\ref{c44}) &  conj.  \tabularnewline
\hline 
$77$  & eq. (\ref{c77}) & conj.  \tabularnewline
\hline 
$78$  & eq. (\ref{c78}) & conj.     \tabularnewline
\hline 
$128$  & eq. (\ref{c128}) &  proof \cite{paper7}\tabularnewline
\hline 
$132$  & eq. (\ref{c132}) & proof \cite{paper7} \tabularnewline
\hline 
$136$  & eq. (\ref{c136}) &  proof \cite{paper7}\tabularnewline
\hline 
$140$  & eq. (\ref{c140}) & proof \cite{paper44}   \tabularnewline
\hline 
$160$  & eq. (\ref{c160}) & proof  \tabularnewline
\hline 
$164$  & unknown &  \tabularnewline
\hline 
$168$  & eq. (\ref{c168}) & proof \tabularnewline
\hline 
$172$  & eq. (\ref{c172})  & conj.  \tabularnewline
\hline 
$232$  & eq. (\ref{c232}) & conj.  \tabularnewline
\hline
\end{tabular}
\end{center}
\caption{Density of ones $c_n$ for arbitrary initial density for elementary rules
asymptotically emulating identity.}
\end{table}
%%%%%%%%%%%%%%%%%%
We have demonstrated that density polynomials are useful for
computing density of ones after $n$ iterations of a CA rule starting
from a Bernoulli distribution. In many CA rules, patterns in density
polynomials can be detected, and then formally proved, such as in the case
of rule 160 and 168. In other cases, one can recognize
in density polynomials known polynomial classes, such as
generalized Lucas polynomials, and then conjecture
closed-form expressions for $c_n$. Our results 
are summarized in Table 2. While at the moment we
do not have  formal proofs of the conjecture formulas,
it is hoped that such proofs can eventually be constructed
using methods similar to those presented here (for rules 160 and 168)
 or in~\cite{paper39}. 
Finally, inspection of Tables 1 and 2 and results we obtained for
rules considered in this paper suggests an interesting possible conjecture.
\begin{conjecture}
 For any CA rule asymptotically emulating identity, the density
of 1's after $n$ iterations, starting from a Bernoulli distribution,
is always in the form
\begin{equation}
 c_n \sim \sum_{i=1}^k a_n \lambda_i^n,
\end{equation}
where $a_i, \lambda_i$ are constants which may only depend on
the initial density $c_0$, and $|\lambda_i|\leq 1$.
\end{conjecture}
Note that some of the $\lambda_i$'s can be complex, and then they come in conjugate pairs,
like in rule 44 (eq. \ref{c44}). When one of the $\lambda_i$'s is equal to 1, then $c_\infty > 0$, otherwise
$c_\infty =0$.

Such behavior of $c_n$ strongly resembles hyperbolicity in finitely-dimensional
dynamical systems.  Hyperbolic fixed points are common type of fixed points in dynamical systems.
If the initial value is near the fixed point and lies on the stable manifold, 
the orbit of the dynamical system converges to the fixed point exponentially fast.
One could argue that the exponential convergence to equilibrium observed
in CA described in this paper is somewhat related to finitely-dimensional hyperbolicity.
We suspect that the the finite-dimensional map,  known as the local structure map~\cite{paper50},
which approximates dynamics of a given CA, should posses a stable hyperbolic fixed point
for every CA asymptotically emulating identity. This hypothesis is currently 
under investigation and will be discussed elsewhere.

\vspace{1em}
\noindent\textbf{Acknowledgments}\\
H. Fuk\'s acknowledges financial support from the Natural Sciences and
Engineering Research Council of Canada (NSERC) in the form of Discovery Grant,
and J.M. G\'omez Soto   acknowledges financial support from Research Council of
M\'exico (CONACYT) and Research Council of Zacatecas (COZYT). 
This work was made possible by the facilities of the Shared
Hierarchical Academic Research Computing Network (SHARCNET:www.sharcnet.ca) and
Compute/Calcul Canada. 

\footnotesize
%\bibliography{emulators,/home/hfuks/ExtData/Bib/hfpub,/home/hfuks/ExtData/Bib/cabooks}
%bibexport.sh -o emulators-ext.bib emulators.aux
\bibliography{emulators-ext.bib}

\end{document}